\documentclass[3p,10pt,times]{elsarticle}

\usepackage{amssymb}
\usepackage{color,colortbl,xcolor}
\usepackage{chemarr}
\usepackage{subfigure} 
\usepackage{graphicx}
\usepackage{epsf,mathtools}
\usepackage[outdir=./]{epstopdf}
\usepackage{epsfig}
\usepackage{dcolumn}
\usepackage{bm}
\biboptions{numbers,sort&compress}
\usepackage{natbib}
\usepackage[british]{babel}
\usepackage{multirow}
\usepackage{multicol}
\usepackage{makecell}

\newcommand{\red}[1]{\textcolor{black}{#1}}

\date{~}

\begin{document}

\begin{frontmatter}



\title{Independent channels for miRNA biosynthesis ensure efficient static and dynamic control in the regulation of the early stages of myogenesis}


\author{Jonathan Fiorentino}
\address{Dipartimento di Fisica, Sapienza Universit\`a di Roma, Rome, Italy}

\author{Andrea De Martino}
\address{Soft \& Living Matter Lab, CNR-NANOTEC, Rome, Italy \\ Italian Institute for Genomic Medicine, Turin, Italy}

\begin{abstract}
Motivated by recent experimental work, we define and study a deterministic model of the complex miRNA-based regulatory circuit that putatively controls the early stage of myogenesis in human. We aim in particular at a quantitative understanding of (i) the roles played by the separate and independent miRNA biosynthesis channels (one involving a miRNA-decoy system regulated by an exogenous controller, the other given by transcription from a distinct genomic locus) that appear to be crucial for the differentiation program, and of (ii) how competition to bind miRNAs can  efficiently control molecular levels in such an interconnected architecture. We show that optimal static control via the miRNA-decoy system constrains kinetic parameters in narrow ranges where the channels are tightly cross-linked. On the other hand, the alternative locus for miRNA transcription can ensure that the fast concentration shifts required by the differentiation program are achieved, specifically via non-linear response of the target to even modest surges in the miRNA transcription rate. While static, competition-mediated regulation can be achieved by the miRNA-decoy system alone, both channels are essential for the circuit's overall functionality, suggesting that that this type of joint control may represent a minimal optimal architecture in different contexts.
\end{abstract}

\end{frontmatter}









\section{Introduction}

Following years of experimental work, microRNAs (miRNAs) --small, endogenous, non-coding RNA molecules ubiquitously found in plant and animal cells-- have emerged as key agents in post-transcriptional regulation \cite{cech2014noncoding}. Their primary mode of action is by protein-mediated base-pairing to target RNA molecules. For coding RNAs, this leads to the repression of gene expression through mRNA cleavage or translational inhibition \cite{hammond2000rna,bartel2004micrornas,bartel2009micrornas,kim2016general}. Their targets however include both coding and non-coding transcripts like long non-coding RNAs (lncRNAs) \cite{ponting2009evolution,rinn2012genome,fatica2014long,engreitz2016long}, which places them at the center of the RNA-based regulatory web. Over time, miRNAs have been found to constitute a remarkably versatile regulatory layer, mediating functions ranging from  the buffering of gene expression noise \cite{siciliano2013mirnas} to the timing of genetic circuits \cite{cheng2007microrna}. Moreover, they are now known to be profoundly implicated in a variety of developmental and disease processes \cite{sayed2011micrornas}.  
Mathematical models have been employed to address the role of miRNAs in different contexts \cite{lai2016understanding}, highlighting for instance how miRNA-based regulation may combine with circuit topology \cite{osella2011role}, kinetic heterogeneities \cite{martirosyan2016probing,marti2} and effects due to competition between miRNA targets \cite{jens2015competition} to generate diverse functional outcomes. Competition, in particular, has been hypothesized to affect regulation in a broader, yet more subtle, way through the so-called `competing endogenous RNA' (ceRNA) effect \cite{salmena2011cerna}, whereby co-regulated targets can establish, under specific conditions \cite{figliuzzi2013micrornas,bosia2013modelling,noorbakhsh2013intrinsic,figliuzzi2014rna,jens2015competition}, an effective crosstalk with potentially far-reaching implications. While experimental validations are currently putting under scrutiny the question of how effective this mechanism can be in standard conditions \cite{bosson2014endogenous,yuan2015model,denzler2016impact}, competition has been shown to be central in a number of situations, perhaps most notably in cancer development \cite{poliseno2010coding,wang2010creb,karreth2015braf} and muscle cell differentiation \cite{cesana2011long,neguembor2014long}. 

Following recent work that has shed new light on its intricate genetic circuitry \cite{legnini2014feedforward}, we focus here on miRNA-based control in early myogenesis. Key actors include two miRNA species (miR-133 and miR-135), two transcription factors (MAML1 and MEF2C), a skeletal muscle-specific lncRNA (linc-MD1) and the RNA-binding human antigen R (HuR) protein (see Fig. \ref{fig:CIRC}A). miR-133 can be produced from a precursor RNA (pre-miR-133b) as well as from two independent genomic loci. However, pre-miR-133b also provides the substrate for the synthesis of linc-MD1 through a pathway alternative (and mutually exclusive) to that leading to miR-133. In addition, linc-MD1 possesses two target sites for miR-135 and one for miR-133 and can therefore act as a `decoy' for both miRNAs. The transcription factors MAML1 and MEF2C, both essential in the expression of muscle-specific genes \cite{shen2006notch}, are instead targets of miR-133 and miR-135, respectively. As a consequence, linc-MD1 is a ceRNA of MAML1 (resp. MEF2C) and competes with it to bind miR-133 (resp. miR-135). In specific, miRNA sponging activity by linc-MD1 de-represses MAML1 and MEF2C leading to muscle-cell differentiation via the expression of the specific genes controlled by the latter. The HuR mRNA plays a subtle role in controlling the alternative processing of pre-miR-133b into linc-MD1 or miR-133. Most importantly, it competes with linc-MD1 for miR-133, thereby favoring the former's sponging activity and giving rise to a positive feedforward loop that ultimately affects the levels of both species. 

As discussed in \cite{legnini2014feedforward}, the trigger that possibly causes the system to exit the feedforward loop, repress muscle-specific gene expression and access later stages of differentiation is an endogenous upregulation of miR-133 transcription from the independent genomic loci. This suggests that the complex regulatory circuitry just described, \red{whereby miRNAs can be synthesized both via a protein controlled switch and from an independent locus}, can provide effective control of both timing and molecular levels. 

In order to analyze this issue in a quantitative framework, here we analyze a schematic version of the above circuitry through a deterministic mathematical model based on mass-action kinetics, focusing specifically on the roles of HuR and of the alternative loci for miRNA transcription. In a nutshell, by characterizing the magnitude of the ceRNA effect and the response to a sudden increase of the transcriptional activity of miR-133, we show, among other things, that, while HuR-controlled regulation of pri-miR-133 processing is crucial to tune molecular levels, miRNA transcription from the alternative loci allows to achieve fast down-regulation of the target transcription factors MAML1 and MEF2C. In particular, fast enough miRNA-ceRNA binding kinetics causes non-linear response of the target level to upshifts in the miRNA biosynthesis rate as moderate as 20\%, providing a highly efficient route to amplifying the effect of the differentiation trigger. In order for both mechanisms to be active, though, kinetic parameters need to be coordinated within \red{specific} ranges of values, which depend strongly on how sensitive pre-miR-133b processing is to HuR levels. In other words, the space of interaction constants and transcription rates is significantly constrained by \red{crosstalk} requirements. 

\red{The fact that crosstalk presupposes some degree of parameter tuning is not surprising {\it per se}, as strong ceRNA-ceRNA effective interactions at stationarity are known to be mainly achieved through competition when the concentrations of the involved molecular species are nearly equimolar \cite{figliuzzi2013micrornas,bosia2013modelling}. Remarkably, though, we find that the system's {\it dynamic} behaviour in such ranges is completely compatible with that observed in time-resolved experiments. This supports the conclusion that the two  regulatory elements of the myogenesis clock, namely the HuR-controlled miRNA-decoy system and the alternative locus for miRNA transcription, play different yet coordinated functional roles.}

\section{Results}

\subsection{Definition of the model}
 
\begin{figure}
\centering\includegraphics[width=13cm]{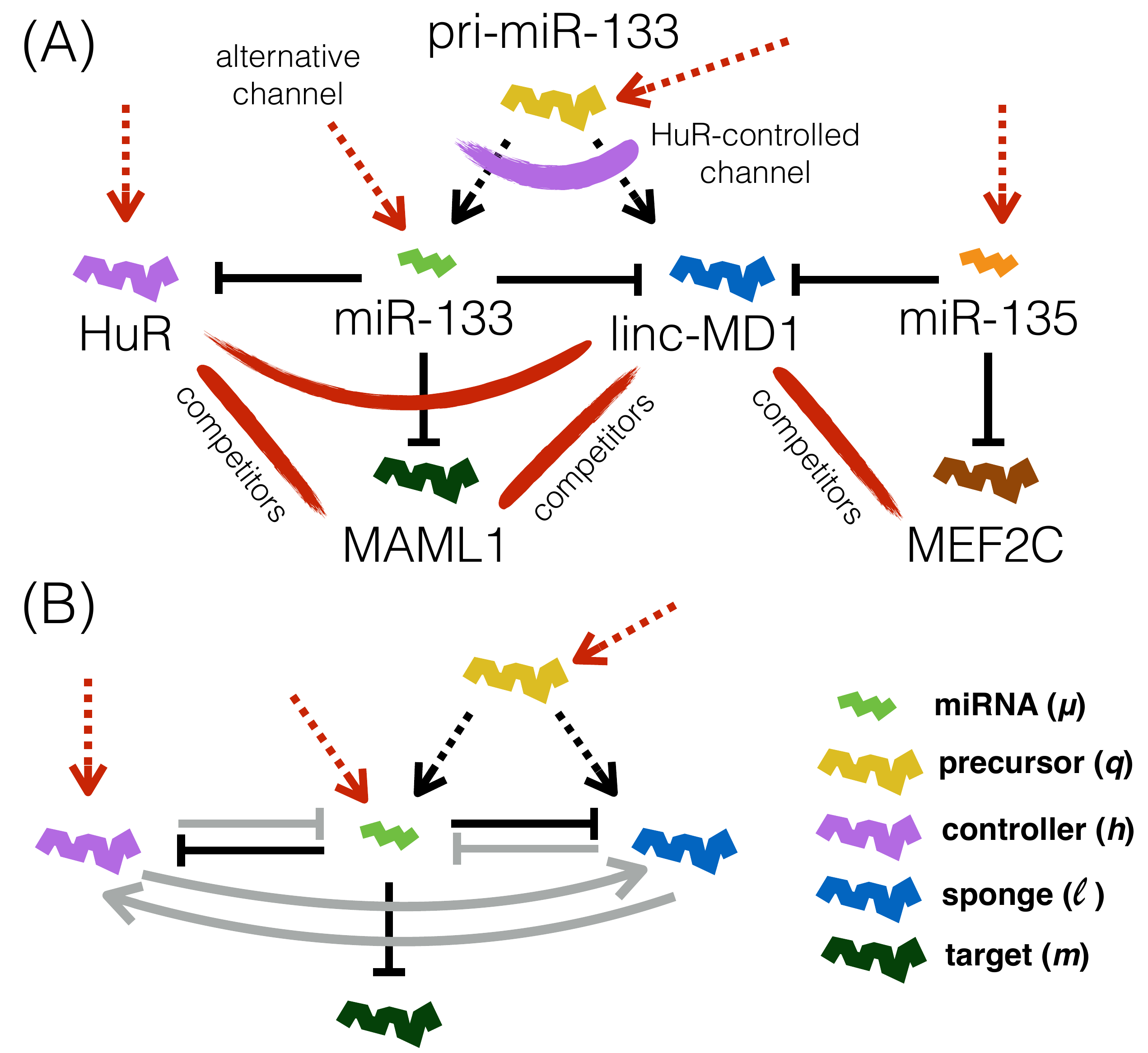}
\caption{Scheme of the miRNA-decoy circuit controlling skeletal muscle-cell differentiation (see text for details). (A) Circuitry identified in \cite{legnini2014feedforward}, showing molecular species and their interactions. Red (resp. black) dashed lines represent standard (resp. mutually alternative) transcriptional modes, while thick red lines indicate that the involved species compete to bind a miRNA. The thick purple line indicates that pri-miR-133 processing is controlled by HuR. (B) Reduced model considered in this paper, with the corresponding nomenclature used in the mathematical model. \red{Grey arrows indicate effective interactions: the sponge and the controller are mutually reinforcing (an increase in the latter leads to increased synthesis of the former, while increased miRNA sponging de-represses the controller), while both effectively repress the miRNA (increased levels of $h$ reduce the synthesis of $\mu$ while increased levels of $\ell$ enhance miRNA sponging).}}\label{fig:CIRC}
\end{figure}

Our model is closely based on the mechanism controlling skeletal muscle cell differentiation identified and discussed in \cite{cesana2011long,legnini2014feedforward} (see Fig. \ref{fig:CIRC}A and B). We consider a precursor RNA species (labeled $q$) which can be processed alternatively into a regulatory microRNA (labeled $\mu$) or a lncRNA (labeled $\ell$). 
The relative weight of the two processing pathways is controlled by another RNA species (labeled $h$), such that larger values of $h$ increasingly favor synthesis of $\ell$ over $\mu$, thereby effectively repressing the latter. $\mu$, however, can directly repress $h$. In addition, $\ell$ can `sponge' (i.e., transiently sequester) $\mu$ and lift its repressive effect from $h$. Finally, $\mu$ negatively controls a target mRNA  (labeled $m$). Ultimately, $\ell$, $m$ and $h$ are ceRNAs, in that they compete to bind $\mu$. Note that, following \cite{legnini2014feedforward}, we allow $\mu$ to be synthesized independently of $q$ from a separate genomic locus. \red{We shall refer to the two distinct miRNA biosynthetic pathways as the `HuR-controlled miRNA channel' and the `alternative miRNA channel', respectively (see Fig. \ref{fig:CIRC}A).} By comparing against the {\it HuR}-controlled miRNA-decoy system of Fig. \ref{fig:CIRC}A, one sees that $q$, $\mu$, $\ell$, $h$ and $m$ play the roles of pri-miR-133b, miR-133, linc-MD1, HuR and MAML1, respectively (we neglected the remaining nodes for sakes of simplicity). 

Our model includes the processes and rates reported in Table \ref{tab:param} together with the values used in this study.
\begin{table}
\centering
\begin{tabular}{lcccc}
\rowcolor{gray!15}
\multicolumn{2}{c}{Processes and rates} & Parameters [units] & In all Figs. & In Fig. \ref{fig:rep}\\
\hline
\multirow{4}{*}{\makecell{Ex-novo synthesis\\ and degradation}} & \multirow{2}{*}{$\emptyset \xrightleftharpoons[d_m]{b_m} m$} & $b_m$ [$\mbox{nM} \cdot \mbox{s}^{-1}$]& $10^{-4}$ & $4\cdot 10^{-5}$\\& & $d_m=d$ [$\mbox{s}^{-1}$] & $10^{-4}$ & $4\cdot 10^{-5}$\\

& \multirow{2}{*}{$\emptyset \xrightleftharpoons[d_q]{b_q} q$} & $b_q$ [$\mbox{nM} \cdot \mbox{s}^{-1}$] & $10^{-4}$ & $6 \cdot 10^{-5}$ \\& & $d_q=d$ [$\mbox{s}^{-1}$] & $10^{-4}$ & $4\cdot 10^{-5}$\\

& \multirow{2}{*}{$\emptyset \xrightleftharpoons[d_h]{b_h} h$} & $b_h$ [$\mbox{nM} \cdot \mbox{s}^{-1}$]& variable & $8\cdot 10^{-5}$\\&  & $d_h=d$ [$\mbox{s}^{-1}$] & $10^{-4}$ & $4\cdot 10^{-5}$\\

& \multirow{2}{*}{$\emptyset \xrightleftharpoons[d_{\mu}]{b_{\mu}} \mu$} & $b_{\mu}$ [$\mbox{nM} \cdot \mbox{s}^{-1}$]& variable & $4 \cdot 10^{-6}$\\& & \red{$d_{\mu}=d$} [$\mbox{s}^{-1}$] & $10^{-4}$ & $4\cdot 10^{-5}$\\

\hline 
\multirow{2}{*}{Synthesis from the precursor} & $q \stackrel{(1-\alpha)b}{\longrightarrow} \mu$ & $\alpha$ [adim.] &  &\\
 & $q \stackrel{\alpha b}{\longrightarrow} \ell$ & $b$ [$\mbox{s}^{-1}$] & $10^{-4}$ & $8 \cdot 10^{-5}$\\

\hline
Degradation of the sponge ($\ell$) & $\ell \stackrel{d_{\ell}}{\longrightarrow} \emptyset$ & $d_\ell=d$ [$\mbox{s}^{-1}$]& $10^{-4}$ & $4\cdot 10^{-5}$\\

\hline

\multirow{3}{*}{\makecell{miRNA-ceRNA complex \\ formation}} & $\mu + m \xrightarrow{k_{\mu m}} c_m$ & $k_{\mu m}$ [$\mbox{nM}^{-1} \cdot \mbox{s}^{-1}$]& $10^{-3}$ & $4\cdot 10^{-4}$\\

& $\mu + \ell \xrightarrow{k_{\mu\ell}} c_{\ell}$ & $k_{\mu\ell}^{\max}$ [$\mbox{nM}^{-1} \cdot \mbox{s}^{-1}$]& $10^{-3}$ & $1.2 \cdot 10^{-4}$\\

& $\mu + h \xrightarrow{k_{\mu h}} c_h$ & $k_{\mu h}$ [$\mbox{nM}^{-1} \cdot \mbox{s}^{-1}$]& $10^{-3}$ & $4\cdot 10^{-4}$\\

\hline

\multirow{3}{*}{\makecell{Stoichiometric complex decay \\ (without recycling of $\mu$)}} & $c_m \stackrel{\sigma_m}{\longrightarrow} \emptyset$ & $\sigma_m=\sigma$ [$\mbox{s}^{-1}$]& $10^{-4}$ & $4\cdot 10^{-5}$\\

& $c_{\ell} \stackrel{\sigma_{\ell}}{\longrightarrow} \emptyset$ & $\sigma_\ell=\sigma$ [$\mbox{s}^{-1}$]& $10^{-4}$ & $4\cdot 10^{-5}$\\

& $c_h \stackrel{\sigma_h}{\longrightarrow} \emptyset$ & $\sigma_h=\sigma$ [$\mbox{s}^{-1}$]& $10^{-4}$ & $4\cdot 10^{-5}$\\

\hline

\multirow{3}{*}{\makecell{Catalytic complex decay \\ (with recycling of $\mu$)}} & $c_m \stackrel{{\kappa}_m}{\longrightarrow} \mu$ & $\kappa_m={\kappa}$ [$\mbox{s}^{-1}$] & $5 \cdot 10^{-4}$ & $2 \cdot 10^{-4}$\\

& $c_{\ell} \stackrel{{\kappa}_{\ell}}{\longrightarrow} \mu$ & $\kappa_\ell={\kappa}$ [$\mbox{s}^{-1}$] & $5 \cdot 10^{-4}$ & $2 \cdot 10^{-4}$\\

& $c_h \stackrel{{\kappa}_h}{\longrightarrow} \mu$ & $\kappa_h={\kappa}$ [$\mbox{s}^{-1}$] & $5 \cdot 10^{-4}$ & $2 \cdot 10^{-4}$\\

\hline

\multirow{2}{*}{Hill indices} & & $n$ [adim.] & 2 & 2\\& & $p$ [adim.] & 2 & 2\\

\hline

\multirow{2}{*}{Dissociation constants} & & $h_{\alpha}$ [$\mbox{nM}$]& $1.0$ & $0.5$\\& & $h_{\mu\ell}$ [$\mbox{nM}$] & $1.0$ & $0.7$\\

\hline

Perturbation fold-size & & $\Delta$ [adim.]& & 5\\  
\hline
\end{tabular}
\caption{Processes included in the model and their associated parameters. The transcription rates $b_h$ (controller) and $b_{\mu}$ (miRNA) are the key control parameters used in this study. Note that $\alpha$ and $k_{\mu\ell}$ are functions of $[h]$, see Eq. (\ref{eqn:alpha}), that vary respectively in $[0,1]$ and $[0,k_{\mu\ell}^{\rm max}]$. The value of $k_{\mu\ell}^{\rm max}$ is as indicated. Numerical values used to obtain Fig. \ref{fig:rep} are shown separately, since a specific fold-size perturbation $\Delta$, describing the increase in $\mu$ biosynthesis due to the alternative channel (see (\ref{pertu})), was chosen in order to reproduce empirical time courses.}  
\label{tab:param}
\end{table}
Note that, since the exact mechanism of target repression by the miRNA is still unclear, we included both a stoichiometric and a catalytic decay mode for miRNA-ceRNA complexes. Indeed, while miRNAs incorporated into the RISC could return to the cytoplasm following target degradation, complexes may be stored in P-bodies, leading to their stoichiometric degradation \cite{valencia2006control}. As said above, $h$ regulates the values of the probability that $q$ is processed into $\ell$ (denoted by $\alpha$) and of the $(\mu,\ell)$ binding rate (denoted by $k_{\mu\ell}$). In \cite{legnini2014feedforward}, it is shown that the HuR protein downregulates the biogenesis of $\mu$ by repressing pri-miRNA cleavage by the enzyme Drosha, fostering accumulation of $\ell$. We therefore assume that $\alpha$ and $k_{\mu\ell}$ are Hill functions of $[h]$ (the level of $h$), with indices $n$ and $p$ respectively, i.e. 
\begin{gather}
\label{eqn:alpha}
\alpha=\frac{[h]^n}{[h]^n+h_{\alpha}^n}~~~~~,~~~~~
k_{\mu\ell}=k_{\mu\ell}^{\max}\frac{[h]^p}{[h]^p+h_{\mu\ell}^p}~~,
\end{gather}
with $h_\alpha$ and $h_{\mu\ell}$ the corresponding dissociation constants. For simplicity, we furthermore neglect miRNA-ceRNA complex dissociation rates, as partially justified by the observation that the ratio between the binding and unbinding rates of complexes is typically large \cite{bosia2013modelling}. Finally, in order to limit parameter proliferation and focus on the action of the HuR-controlled switch and of the alternative miRNA channel, we assume homogeneous degradation and complex processing constants for RNA molecules. \red{(The point where biological realism is most sacrificed by this choice lies in the assumption that the intrinsic decay rates for miRNAs and ceRNAs are comparable whereas such rates are most likely linked to RNA length. This is however only going to affect our results in a quantitative way.)}

With these choices, the mass-action dynamics of the regulatory element shown in Fig. \ref{fig:CIRC}B is described by 
\begin{gather}
\label{eqn:mev}
\frac{d}{dt}[m]=b_m-(d+k_{\mu m}[\mu])[m]~~,\\
\frac{d}{dt}[\mu]=b_{\mu}+(1-\alpha )b[q]+\kappa([c_{\ell}]+[c_m]+[c_h])-(d_{\mu}+k_{\mu m}[m]+k_{\mu\ell}[\ell]+k_{\mu h}[h])[\mu]~~,\\
\label{eqn:lev}
\frac{d}{dt}[\ell]=\alpha b[q]-(d+k_{\mu\ell}[\mu])[\ell]~~,\\
\frac{d}{dt}[h]=b_h-(d+k_{\mu h}[\mu])[h]~~,\\
\frac{d}{dt}[q]=b_q-(d+b)[q]~~,\\
\frac{d}{dt}[c_m]=-(\sigma + \kappa)[c_m]+k_{\mu m}[\mu][m]~~,\\
\frac{d}{dt}[c_{\ell}]=-(\sigma+ \kappa)[c_{\ell}]+k_{\mu\ell}[\mu][\ell]~~,\\
\label{eqn:chev}
\frac{d}{dt}[c_h]=-(\sigma + \kappa)[c_h]+k_{\mu h}[\mu][h]~~.
\end{gather}

\subsection{Comparison with experimental time courses and choice of parameters}

Time courses for the levels of linc-MD1 ($\ell$ in our model), miR-133 ($\mu$) and HuR ($h$) for a mouse myoblast cell culture capable of differentiation have been characterized in \cite{legnini2014feedforward}. A Western blot showing the behaviour of MAML1 ($m$) during \textit{in vitro} differentiation is instead displayed in \cite{cesana2011long}. These results indicate that 24 hours after the induction of differentiation the levels of $\ell$ and $\mu$ are low, while the controller \red{$h$} is abundant. At this stage, $\mu$ is only synthesized from the precursor $q$ (i.e. from the genomic locus miR-133b). On the other hand, $[m]$ is at its maximum. In this phase, myoblasts differentiate and  consistently express the target. 48 hours after induction, the level of $h$ peaks while that of the lncRNA $\ell$ appears to be increasing. This continues until about 72 hours after induction, during which time frame $[\mu]$ also increases thanks to a rapid increase in miRNA synthesis from the independent genomic locus miR-133a \cite{legnini2014feedforward}. In this scenario, $\ell$ can sponge miRNAs away from the target $m$. When myoblast differentiation is accomplished (at about 72 hours after induction), though, $m$ must be repressed. Because $[h]$, and therefore $[\ell]$, appear to be decreasing between 72 and 96 hours, the sponge $\ell$ is no longer able to efficiently remove miRNAs from the target, which therefore gets silenced. This brings the cell to the next stage of differentiation, namely the formation of myocytes, where the expression of a different set of  genes is paramount \cite{fatica2014long}.

One sees that (i) $\mu$ is the agent that represses the target to complete the differentiation stage, (ii) $\ell$ plays a key role in controlling the timing of differentiation when the miRNA population increases following the activation of the alternative locus, while (iii) the controller ($h$) ensures that $\ell$ is synthesized at sufficiently high rates at the expense of $\mu$ via the $q$-processing switch.
  
\begin{figure}
\centering
\includegraphics[width=14cm]{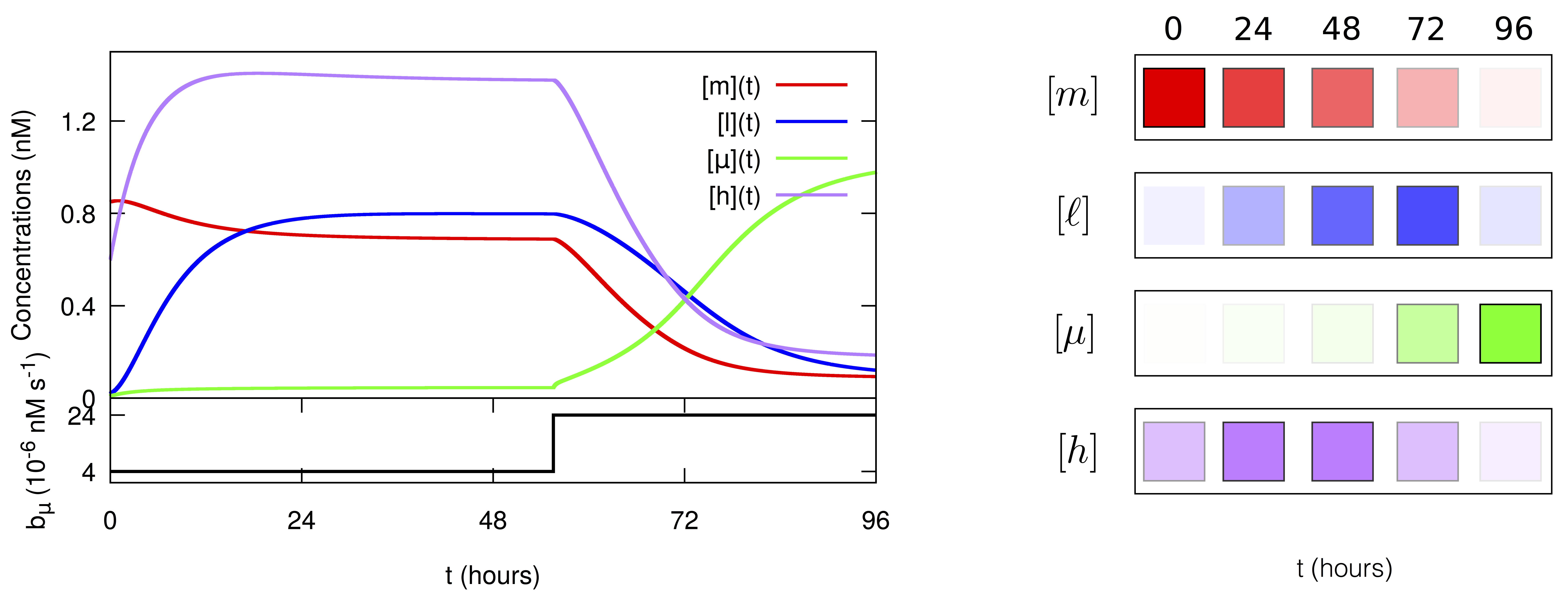}
\caption{Dynamical behaviour of the model, to be compared with the experimental time courses presented in \cite{legnini2014feedforward,cesana2011long} \red{and reported schematically on the right}. The controller ($h$) drives the synthesis of the sponge ($\ell$) at rates high enough to efficiently sequester miRNAs ($\mu$) until the transcription from the alternative locus markedly increases at $t^{\star} \simeq 50$ h. In turn, $\ell$ ensures that the target stays de-repressed long enough to bring the differentiation process to maturation. When the level of $\mu$ becomes sufficiently high, $m$ is rapidly silenced and the next stage of differentiation sets in.}\label{fig:rep}
\end{figure}

We have modeled the increase in miR-133 production from the independent genomic locus through a step-wise perturbation on $b_{\mu}$ of the form
\begin{equation}\label{pertu}
b_{\mu}(t)= b_{\mu} \left[1+\Delta \theta(t-t^{\star})\right]~~,
\end{equation}
where $\Delta$ measures the fold change in transcription rate  after time $t^{\star}$ and $\theta(x)$ is the Heavyside step function. Using (\ref{pertu}) together with the values of the kinetic parameters listed in Table \ref{tab:param}, we obtain the scenario displayed in Figure \ref{fig:rep}, in which the concentrations of $h$, $m$, $\ell$ and $\mu$ are shown as functions of  time. 
Prior to the perturbation, which takes place at $t^\star\simeq 50$ h, the miRNA level is small while $[h]$ and $[\ell]$ rise together as the target level $[m]$ decreases. When the alternative channel for miRNA transcription sets in, $\mu$ levels shoot up, thereby repressing $m$, $\ell$ and $h$, in \red{reasonable} qualitative (for concentrations) and quantitative (for time scales) agreement with observations \cite{legnini2014feedforward}. (Note that results have been obtained by fixing $\Delta=5$.)

For the remaining analysis, the choice of kinetic parameters was based on loose ranges extrapolated from \cite{wang2010toward,haley2004kinetic,alon2006introduction,palssonsystems}, namely 
\begin{gather}
d \simeq \frac{1}{\tau}~~~,~~~\sigma \simeq \frac{1}{\tau}~~~,~~~b_j \simeq \frac{\gamma}{\tau}~~,\\
k_{\mu i} \in \left[\frac{1}{100\gamma\tau},\frac{100}{\gamma\tau}\right]~~~,~~~
\kappa \in \left[\frac{1}{10\tau},\frac{10}{\tau}\right]~~,
\end{gather}
where $j\in\{h,m,\mu,q\}$, $i\in\{h,\ell,m\}$, $\tau \simeq 10^{4}\; \mbox{s}$ is a typical mRNA half life in human cells and $\gamma \simeq 1\;\mbox{nM}$ a reference RNA concentration. The values of the kinetic parameters used in the numerical study of the steady state are shown in Table \ref{tab:param}.

\subsection{ceRNA crosstalk at steady state}

\begin{figure}
\centering
\includegraphics[width=\textwidth]{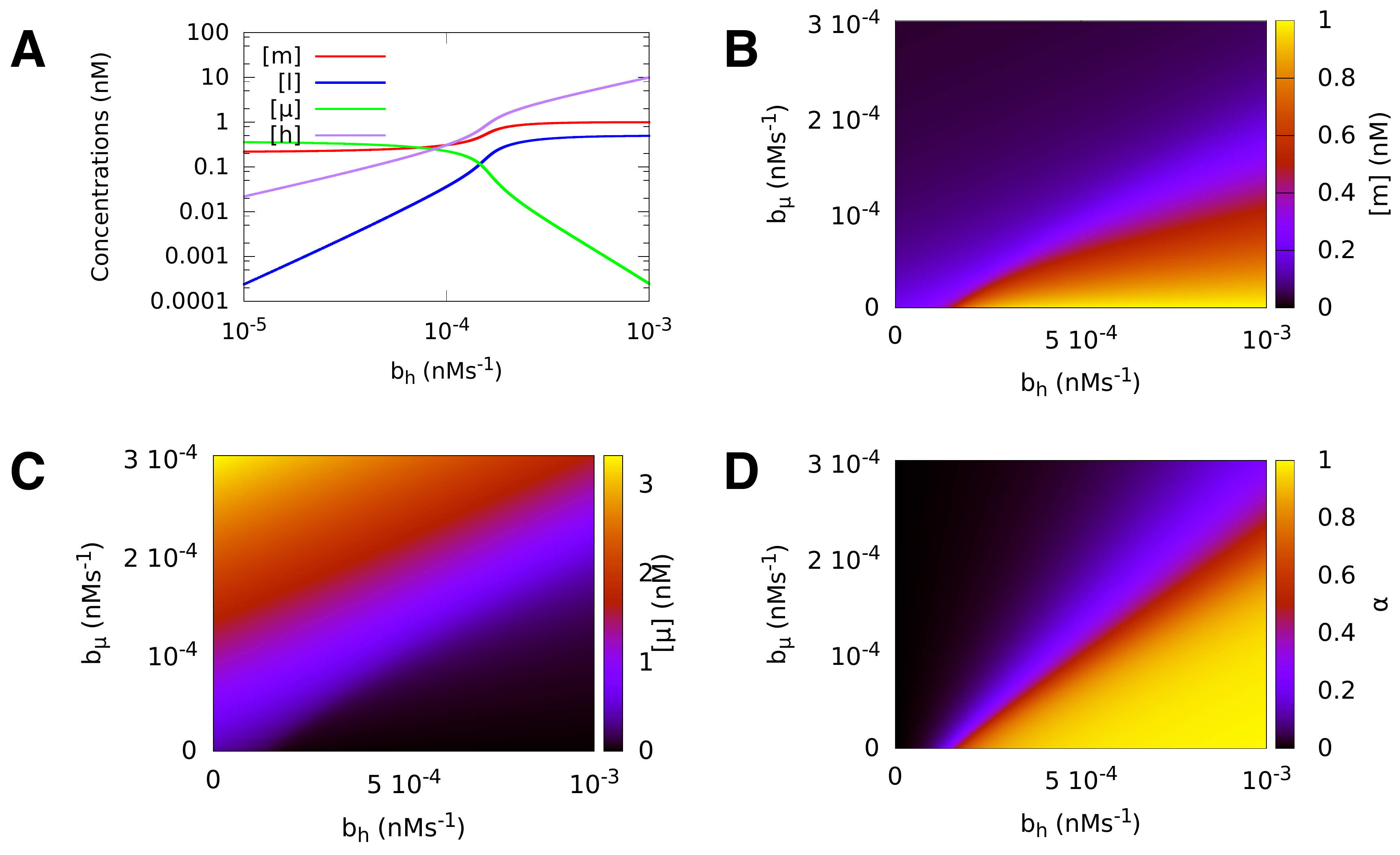}
\caption{(A) Steady state concentrations of $m$, $\ell$, $h$ and $\mu$ as functions of $b_h$ with all other parameters kept constant and $b_{\mu}=0$. A change in the transcription rate of $h$ affects the levels of $m$ and $\ell$, which crosstalk through $\mu$. (B--D) Heat maps of $[m]$, $[\mu]$ and $\alpha$ returning their values versus $b_h$ and $b_\mu$. As one would expect, the behaviour of $[\mu]$ is roughly opposite to that of $[m]$. Notice that in each case the dynamical range, where molecular levels are maximally sensitive to changes in the parameters, occurs for similar values of $b_h$ and $b_\mu$.} 
\label{fig:steady}
\end{figure}

We shall now quantitatively assess the ability of the miRNA-decoy system to control the expression of the target by analyzing both steady state and dynamical responses to perturbations. We focus specifically on the miRNA-mediated ceRNA mechanism, whose regulatory effectiveness has been quantified in \cite{martirosyan2016probing}. Molecular levels at steady state are easily seen to satisfy the conditions
\begin{gather}
\label{eqn:muss}
[\mu]=\frac{b_{\mu}+(1-\alpha)b[q]+\kappa([c_{\ell}]+[c_m]+[c_h])}{d_{\mu}+k_{\mu m}[m]+k_{\mu\ell}[\ell]+k_{\mu h}[h]}~~,\\
\label{eqn:mss}
[m]=\frac{b_m}{d+k_{\mu m}[\mu]}~~~,~~~
[\ell]=\frac{\alpha b[q]}{d+k_{\mu\ell}[\mu]}~~,~~~
[h]=\frac{b_h}{d+k_{\mu h}[\mu]}~~~,~~~
[q]=\frac{b_q}{d+b}~~,\\
\label{eqn:cmss}
[c_m]=\frac{k_{\mu m}[\mu][m]}{\sigma+\kappa}~~~,~~~
[c_{\ell}]=\frac{k_{\mu\ell}[\mu][\ell]}{\sigma+\kappa}~~~,~~~
[c_h]=\frac{k_{\mu h}[\mu][h]}{\sigma+\kappa}~~.
\end{gather}
Solutions (see Fig. \ref{fig:steady}A)  show that a change in the transcription rate of HuR ($h$, the controller) with all other parameters fixed and no miRNA transcription from the alternative site affects the steady state concentrations of all other ceRNAs. For small enough $b_h$, the repressive effect of $\mu$ on the other RNAs is dominant, so the target ($m$) attains a low expression level. As $b_h$ increases, $[h]$ and, in turn, $\alpha$ and $k_{\mu\ell}$ increase according to (\ref{eqn:alpha}), enhancing the synthesis of the lncRNA $\ell$ (the sponge), which gradually de-represses the target by sequestering miRNAs. As a consequence, the target level increases. For large enough $b_h$, $m$ and $\ell$ attain finite values while $h$ increases linearly (as expected) and the level of free miRNAs gets more and more suppressed. 

Overall, Fig. \ref{fig:steady}A describes how competition for $\mu$ controls the expression levels of the competitors ($m$, $\ell$ and $h$) through modulation of the rate of synthesis of the controller and via the action of the $q$-processing switch alone (as $b_\mu$ is taken to be zero). Note that the regulatory element appears to be more sensitive to changes in $b_h$ in a relatively narrow range of values. In this range, molecular populations achieve near equimolarity, a condition under which the magnitude of miRNA-mediated crosstalk is maximal \cite{bosia2013modelling,figliuzzi2013micrornas}. 

Figures \ref{fig:steady}B--D display the dependence of $[m]$, $[\mu]$ and $\alpha$ on $b_h$ and on the rate $b_\mu$ of exogenous miRNA transcription, i.e. from the independent genomic locus miR-133a.
As $b_\mu$ increases, the window of values of $b_h$ where the ceRNA effect is stronger shifts towards higher values. Expectedly, though, the magnitude of target de-repression decreases significantly as more miRNAs become available (Fig. \ref{fig:steady}B), implying that effective ceRNA crosstalk requires a fine tuning of transcription levels from the alternative miRNA locus. Likewise, \red{small values of $b_h$ lock the precursor processing pathway into synthesizing $\mu$ preferentially over the sponge $\ell$ (i.e. $\alpha\simeq 0$), see Fig. \ref{fig:steady}D.} Note that the transition between the two states of the HuR-controlled switch is sharper when the transcriptional activity from the independent miRNA channel is low and becomes more graded for high values of $b_\mu$. \red{This suggests that, in order to avoid step-wise transitions (which would impose a high extra metabolic burden on cells and could be affected by different types of noise, including cell-to-cell variability in transcription rates) as well as slow cross-overs (which would require large changes in transcriptional activity to robustly switch from one state to the other), both $b_\mu$ and $b_h$ should lie in an intermediate range of values.}

\subsection{Optimal control is achieved in specific ranges of kinetic parameters}

\red{In order to quantify the magnitude of ceRNA crosstalk and to understand how crosstalk requirements constrain kinetic parameters, we follow \cite{figliuzzi2013micrornas,figliuzzi2014rna} and focus on the  `susceptibilities'}
\begin{gather}
\chi_{mh}=\frac{\partial{[m]}}{\partial{[h]}}\label{chimh}\\
\chi_{mb_{\mu}}=\frac{\partial{[m]}}{\partial{b_{\mu}}}~~,\label{chimbmu}
\end{gather}
which represent, respectively, the response of the target $m$ to a (sufficiently small) variation in the level of the controller $h$ or the rate of miRNA transcription from the alternative channel. \red{Such quantities, similar to those employed for the study of magnetic spin systems in statistical physics, are especially useful for the analysis of ceRNA networks as they allow to focus specifically on the effects induced by competition to bind miRNAs. In principle, the existence of ceRNA crosstalk could also be signaled by a positive (Pearson) correlation between ceRNA levels. However, positive correlations can simply result from random fluctuations of miRNA levels: as both targets respond to changes in miRNA availability, their fluctuations are expected to be correlated when miRNA levels fluctuate stochastically. This however does not necessarily imply that the level of a ceRNA will change when the level of its competitor is altered. Quantities like (\ref{chimh}), on the other hand, can be non-zero even for a deterministic system and therefore capture precisely the competition-induced aspect highlighted by the experimental literature. (See \cite{translating} for a detailed study of the relationship between susceptibilities and correlations in ceRNA networks.)}

In order to calculate $\chi_{mb_{\mu}}$, we notice that  \eqref{eqn:muss} can be re-written as
\begin{equation}
\label{eqn:musost}
[\mu]=\frac{b_{\mu}+(1-\alpha)b[q]}{d_{\mu}+w(k_{\mu m}[m]+k_{\mu\ell}[\ell]+k_{\mu h}[h])}~~,
\end{equation}
where
\begin{equation}
w=\frac{\sigma}{\sigma+\kappa}
\end{equation}
is the stoichiometricity ratio quantifying the relative weight of the stoichiometric pathway of miRNA-ceRNA complex degradation. In turn, $[m]$, $[\ell]$ and $[h]$ take the form
\begin{gather}
\label{eqn:msost}
[m]=\frac{b_m}{d+k_{\mu m}\,[\mu]}~~~,~~~
[\ell]=\frac{\alpha b[q]}{d+k_{\mu\ell}\,[\mu]}~~~,~~~
[h]=\frac{b_h}{d+k_{\mu h} \,[\mu]}~~.
\end{gather}
From (\ref{eqn:musost}) and (\ref{eqn:msost}) one derives a system of three equations for the susceptibilities $\chi_{mb_{\mu}}$, $\chi_{\ell b_{\mu}}$, $\chi_{hb_{\mu}}$, namely 
\begin{gather}
\chi_{mb_{\mu}}=\frac{[\mu] [m]^2k_{\mu m}}{b_m}\,G(\boldsymbol{\chi})~~,\\
\label{chilbmu}
\chi_{{\ell}b_{\mu}}=\frac{[\mu] [{\ell}]^2k_{\mu\ell}}{\alpha b[q]}\,G(\boldsymbol{\chi})+\frac{[\ell]}{\alpha}\left(\frac{d\alpha}{d[h]}-\frac{[\ell][\mu]}{b[q]}\frac{dk_{\mu\ell}}{d[h]}\right)\chi_{hb_{\mu}}~~,\\
\chi_{hb_{\mu}}=\frac{[\mu] [h]^2k_{\mu h}}{b_h}\,G(\boldsymbol{\chi})~~,
\end{gather}
with $\boldsymbol{\chi}=\{\chi_{mb_{\mu}},\chi_{\ell b_{\mu}},\chi_{hb_{\mu}}\}$ and
\begin{equation}
\label{G_chi}
G(\boldsymbol{\chi})=\frac{w[\mu]\left[k_{\mu m}\chi_{mb_{\mu}}+k_{\mu\ell}\chi_{\ell b_{\mu}}+\left(k_{\mu h}+\frac{b[q]}{[\mu]w}\frac{d\alpha}{d[h]}+[\ell]\frac{dk_{\mu\ell}}{d[h]}\right)\chi_{hb_{\mu}}\right]-1}{b_{\mu}+(1-\alpha)b[q]}~~,
\end{equation}
where
\begin{gather} 
\frac{d\alpha}{d[h]}=\frac{nh_{\alpha}^n[h]^{n-1}}{([h]^n+h_{\alpha}^n)^2}~~,\\
\frac{dk_{\mu\ell}}{d[h]}=k_{\mu\ell}^{\max}\frac{ph_{\mu\ell}^p[h]^{p-1}}{([h]^p+h_{\mu\ell}^p)^2}~~.
\end{gather}
Upon solving the above system one obtains explicit expressions for the susceptibility vector $\boldsymbol{\chi}$. Specifically, for $\chi_{mb_\mu}$, Eq. (\ref{chimbmu}), one gets
\begin{gather}
\chi_{mb_{\mu}}=\frac{k_{\mu m} b_h [\mu][m]^2}{w[\mu]^2\mathcal{A}-b_m b_h \Big[b_{\mu}+(1-\alpha)b[q]\Big]}~~, \label{eqn:chimbmu}\\
\mathcal{A}=k_{\mu \ell}^2 [\ell]^2  \frac{b_m b_h}{\alpha b [q]}+k_{\mu m}^2[m]^2b_h+k_{\mu h}[h]^2b_m \left[ k_{\mu h}+ \left(\frac{k_{\mu \ell}[\ell]}{\alpha}+\frac{b[q]}{[\mu]w}\right)\frac{d \alpha}{d [h]}+\left(1-\frac{[\ell][\mu]k_{\mu \ell}}{\alpha b [q]}\right)[\ell]\frac{d k_{\mu \ell}}{d [h]}  \right]~~\nonumber.
\end{gather}
By similar steps, one can compute susceptibilities such as $\chi_{m h}$, Eq. (\ref{chimh}), for which we find in particular
\begin{multline}
\label{eqn:chimh}
\chi_{mh}= \frac{[m]^2k_{\mu m}[\mu] \alpha b [q]}{\alpha b[q]b_m[b_{\mu}+(1-\alpha)b[q]]-w[\mu]^2({k_{\mu m}}^2[m]^2\alpha b[q]+{k_{\mu\ell}}^2[\ell]^2b_m)}\times\\
\times\left\{b[q]\frac{d\alpha}{d[h]}+w[\mu]\left[k_{\mu h} +\frac{dk_{\mu\ell}}{d[h]}\ell+\frac{k_{\mu\ell}[\ell]}{\alpha}\left(\frac{d\alpha}{d[h]}-\frac{[\ell][\mu]}{b[q]}\frac{dk_{\mu\ell}}{d[h]}\right)\right]\right\}.
\end{multline}

\begin{figure}
\centering
\includegraphics[width=\textwidth]{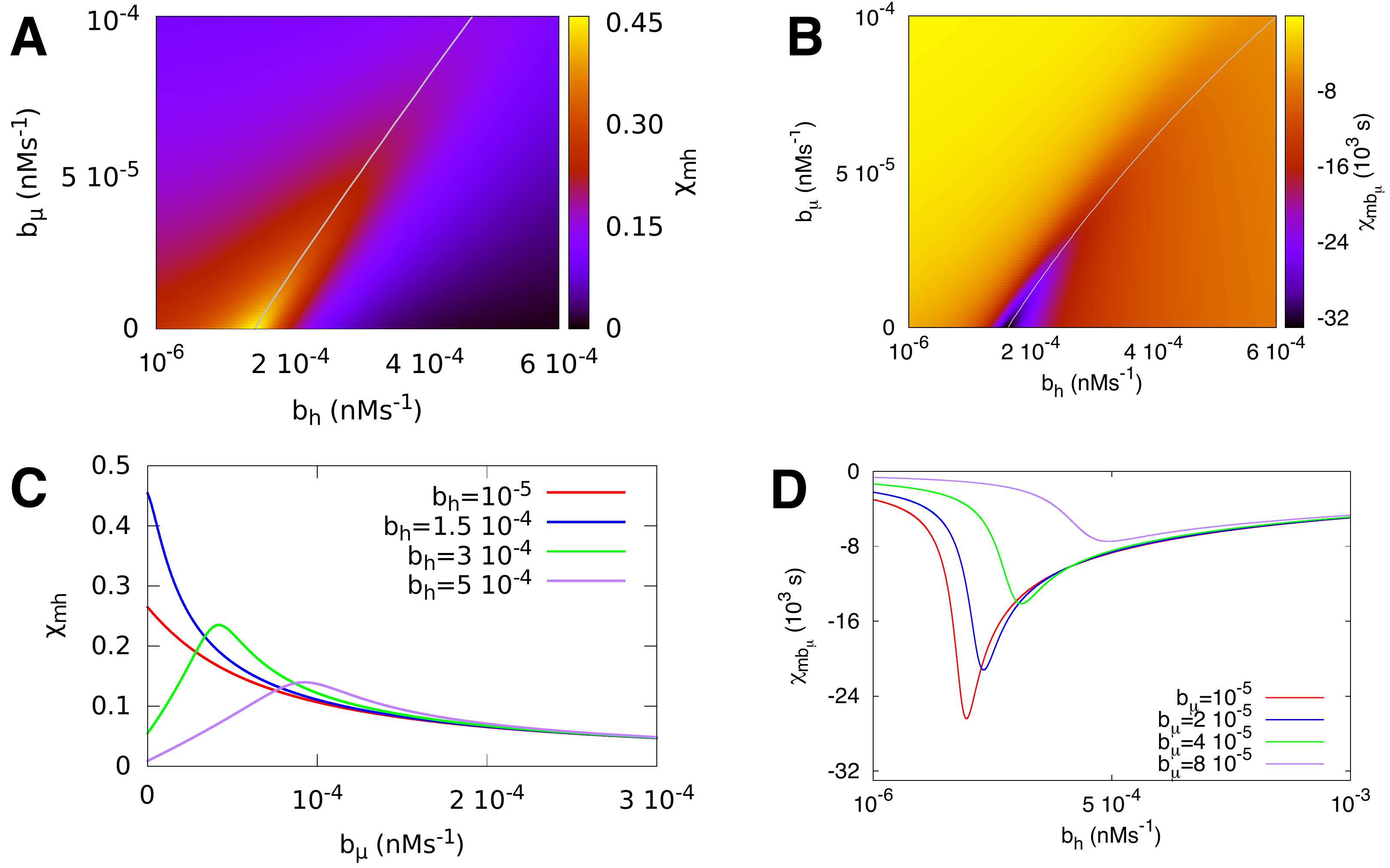}
\caption{(A--B) Heat maps of $\chi_{mh}$ and $\chi_{mb_{\mu}}$ as functions of $b_h$ and $b_{\mu}$. The grey lines represent,  respectively, the curve of maximum $\chi_{mh}$ and minimum $\chi_{mb_{\mu}}$, where control is optimized at fixed $b_h$. In both cases, optimal target control at stationarity is achieved for $b_h\simeq 2\cdot 10^{-4}\; \mbox{nM} \cdot \mbox{s}^{-1}$ and very small $b_\mu$. (C) Indeed, for $b_h=10^{-5}\; \mbox{nM} \cdot \mbox{s}^{-1}$ and $b_h=1.5 \cdot 10^{-4}\; \mbox{nM} \cdot \mbox{s}^{-1}$ (red and blue curves), $\chi_{mh}$ is maximum for $b_{\mu}=0$ and it is monotone decreasing with $b_\mu$. This behaviour however changes drastically for slightly larger $b_h$, when $\chi_{mh}$ peaks at a non-zero value of $b_\mu$. (D) Likewise, optimal $\chi_{mb_{\mu}}$ is always achieved for $b_h \neq 0$, although its absolute value gets smaller as $b_{\mu}$ increases.}
\label{fig:susc}
\end{figure}

The heat maps displayed in Fig. \ref{fig:susc}A and B represent $\chi_{mh}$ and $\chi_{mb_{\mu}}$ as functions of $b_h$ and $b_\mu$, with grey lines showing the curve of maximum $\chi_{mh}$ and minimum $\chi_{mb_{\mu}}$, respectively. (Note that while $\chi_{mh}$ is positive or zero since $h$ represses $\mu$ that in turn represses $m$, $\chi_{mb_\mu}$ is negative or zero, since a positive shift of $b_{\mu}$ causes the level of $m$ to decrease.) The key observation is that the miRNA-decoy system appears to achieve optimal target control in a restricted range of kinetic parameters. 

Figure \ref{fig:susc}C and D detail these effects with a higher resolution. From the former one sees that for $b_h=10^{-5}\; \mbox{nM} \cdot \mbox{s}^{-1}$ (red curve) $\chi_{mh}$ is maximum for $b_{\mu}=0$ and it is monotone. In other terms, for a sufficiently small $b_h$, i.e. when the expression level of the controller is  low enough, the target level appears to be maximally sensitive to the controller level when the alternative miRNA channel is inactive. As the controller's level increases, instead, optimal control requires a non-zero transcriptional activity from the alternative miRNA locus. A similar scenario is obtained by analyzing the sensitivity of the target to $b_\mu$ upon changing $b_h$. Overall, optimal control is achieved in a restricted range of parameters characterized by $b_h\simeq 2.5\cdot 10^{-4}\; \mbox{nM} \cdot \mbox{s}^{-1}$ and small $b_\mu\gtrsim 10^{-5}\; \mbox{nM} \cdot \mbox{s}^{-1}$.

To sum up, we see that effective target control via $h$ can be achieved at steady state in absence of the alternative miRNA channel. However, if both the HuR-controlled channel and the alternative miRNA channel are active, then optimal regulation requires that their rates are coordinated (the faster the former, the faster the latter). On the other hand, effective target control by the alternative miRNA channel requires a controller. This provides quantitative support to the idea that the controller ($h$) ensures fine tuning at stationarity, when the alternative miRNA channel is off, while the latter appears to serve the mainly dynamical purpose of guaranteeing fast target downshifts, in full agreement with results obtained in \cite{figliuzzi2014rna}, where a dynamical analysis of the ceRNA effect has shown that the optimal strategy to achieve fast target downregulation consists in rapidly upregulating miRNA levels.

\subsection{The HuR-controlled channel dominates miRNA biosynthesis when ceRNA crosstalk is optimal}

\begin{figure}
\centering
\includegraphics[width=\textwidth]{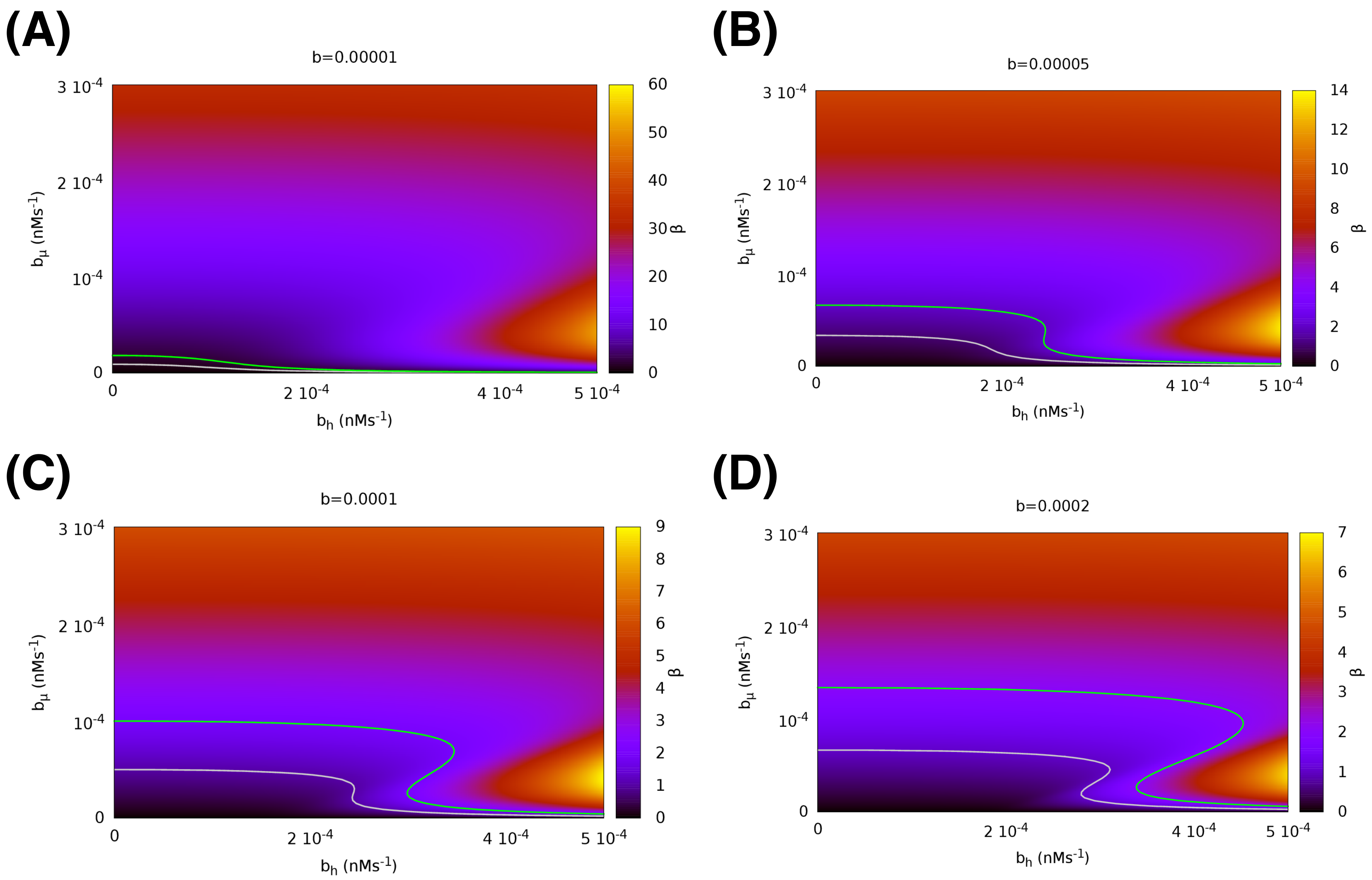}
\caption{Heat maps showing the value of $\beta$, Eq. \eqref{eqn:beta}, as a function of $b_h$ and $b_\mu$ for different values of $b$ (increasing, as displayed, from (A) to (D)). The grey (resp. green) curve gives the $\beta=1$ (resp. $\beta=2$) contour.}\label{fig:fisio}
\end{figure}

In order to further clarify how the interplay between the two modes of miRNA biosynthesis impacts ceRNA crosstalk, we analyze here how the adimensional quantity
\begin{equation}
\label{eqn:beta}
\beta\equiv\frac{b_{\mu}}{(1-\alpha) b[q]}=\frac{b_{\mu}([h]^n+h_{\alpha}^n)}{b[q]h_{\alpha}^n}~~,
\end{equation}
representing the ratio between the miRNA transcription rates from the independent genomic locus ($b_\mu$, corresponding to miR-133a) and from the precursor (corresponding to miR-133b), changes upon changing $b_\mu$ and the transcription rate of the controller, $b_h$. Figure \ref{fig:fisio} displays results, with the grey and green curves outlining, respectively,  the $\beta=1$ and $\beta=2$ contours as a reference.
We see that the HuR-controlled channel is consistently dominant for small enough $b_\mu$ and $b_h$ (where ceRNA control is most efficient), while the alternative channel becomes increasingly more relevant as transcription rates increase. This again suggests that activation of transcription from the independent locus mainly plays a role in determining fast target downregulation. Notice that a two-fold increase of transcription rates with respect to optimal values for ceRNA crosstalk would be required in order to substantially alter this scenario.

\subsection{The Hill index characterizing precursor processing sharpens the region of maximum ceRNA crosstalk}

\begin{figure}
\centering
\includegraphics[width=\textwidth]{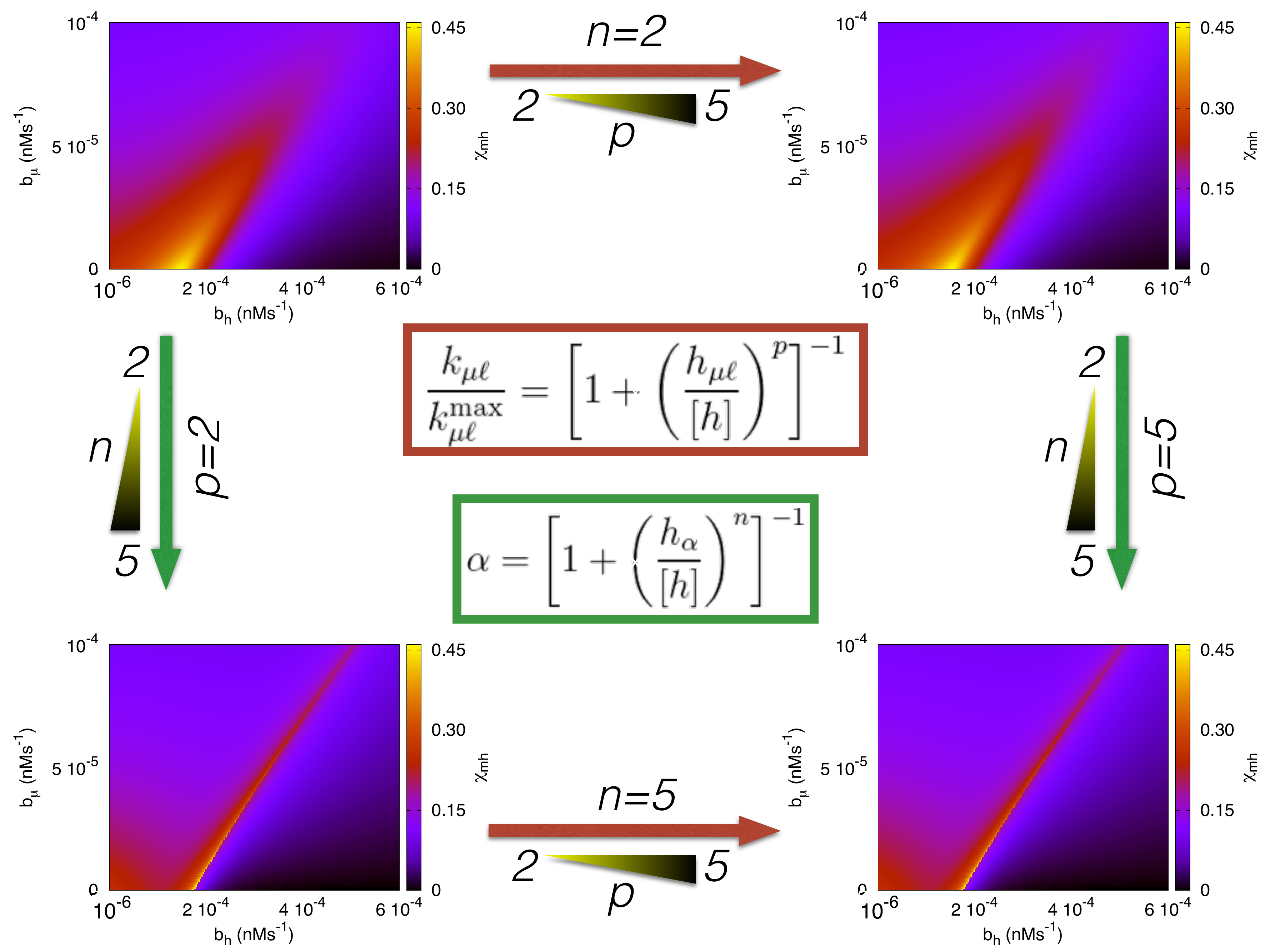}
\caption{Behaviour of the susceptibility $\chi_{mh}$ versus $b_h$ and $b_\mu$ upon varying the Hill indices $n$ (controlling the steepness of the dependence of $\alpha$ on the controller level $[h]$) and $p$ (controlling the steepness of the dependence of $k_{\mu\ell}$ on $[h]$). Top left panel: $n=p=2$. Top right: $n=2$, $p=5$. Bottom left: $n=5$, $p=2$. Bottom right: $n=p=5$. $\chi_{mh}$ is only weakly affected by changes in $p$, while it is very sensitive to $n$. In particular, the optimal range of values of $b_h$ (for fixed $b_\mu$) generically shrinks, while a more tight control can be obtained for larger values of $b_\mu$.}
\label{fig:robust}
\end{figure}

In Eq. \eqref{eqn:alpha}, we have assumed that the parameters $\alpha$ and $k_{\mu\ell}$ are Hill functions of the HuR level $[h]$, and the corresponding Hill indices $n$ and $p$ have been so far kept fixed (see Table \ref{tab:param}). Interestingly, though, they appear to have different effects on the emerging crosstalk scenario (see Fig. \ref{fig:robust}).
An increase of the Hill index $p$ controlling the sharpness of $k_{\mu\ell}$'s response to $[h]$ does not significantly influence the behaviour of the susceptibility $\chi_{mh}$. On the other hand, by increasing $n$ (which modulates $\alpha$'s response to $[h]$), the region of maximal susceptibility gets much sharper while preserving its overall qualitative behaviour. Notice that a larger $n$ would allow to stretch optimal control to much higher values of $b_\mu$, albeit at increased metabolic  costs associated to the fine-tuning of $b_\mu$ and $b_h$ and without significant gains in terms of $\chi_{mh}$. Whether an increased sensitivity of the miRNA-decoy switch is advantageous therefore appears to depend crucially on the transcriptional activity of the alternative miRNA channel, with lower (resp. higher) levels requiring lower (resp. higher) sensitivity for $\alpha$.

\subsection{Fast binding kinetics leads to non-linear response}

\begin{figure}
\centering
\includegraphics[width=9cm]{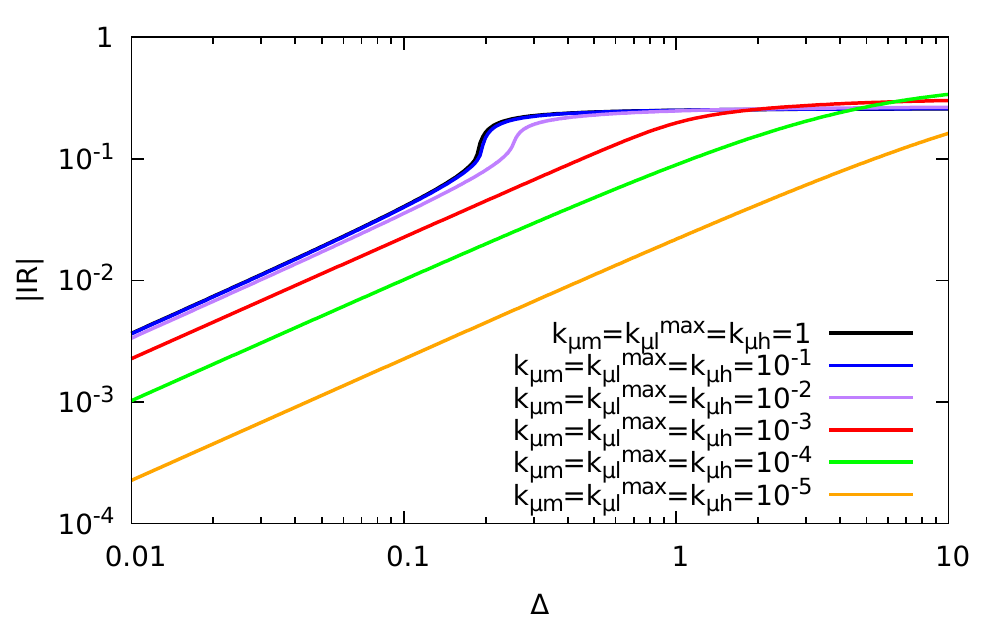}
\caption{Integrated Response IR (absolute value) as a function of the fold-increase $\Delta$ of the transcriptional activity from the alternative locus for miRNA expression, for several values of the miRNA-ceRNA binding rates. Generically, stronger coupling constants lead to non-linear target response already for $\Delta\simeq 20$\%, while the response is linear for smaller binding rates. Results obtained for $b_h = 5\cdot 10^{-4}$ and $b_\mu = 5\cdot 10^{-5}$.}
\label{fig:intres}
\end{figure}

The effectiveness of the mechanism allowing to exit the myoblast stage of differentiation can be analyzed by studying the change in the steady-state value of $m$ as a function of the perturbation size $\Delta$ (see Eq. \eqref{pertu}), quantified by the Integrated Response \cite{figliuzzi2014rna} 
\begin{equation}
\mathrm{IR}(\Delta)= \int_0^{\infty} \Big[[m](t+t^{\star})-[m](t^{\star})\Big]dt~~,
\end{equation}
where $t^\star$ stands for the time at which the perturbation is switched on. $\mathrm{IR}$ clearly depends on $\Delta$, \red{as well as on the duration of the perturbation, and it is sensitive to the subsequent relaxational dynamics. In the context of miRNA-ceRNA networks, it has the extra advantage that it is directly linked to the total amount of protein synthesized from $m$ in response to the perturbation, which represents the key output variable in transcriptional regulatory systems.} Note that, upon activating the alternative miRNA transcriptional channel with rate $b_\mu$, $[m]$ decreases monotonically, so that $\mathrm{IR}$ is always negative. Numerical results obtained for different values of the perturbation size $\Delta$ and for several values of the binding rates $k_{\mu m}$, $k_{\mu\ell}^{\max}$ and $k_{\mu h}$  are shown in Fig. \ref{fig:intres}.
For small values of the binding rates (see e.g. orange and green curves) $\mathrm{IR}$ is roughly linear in $\Delta$. As the binding rates increase, though, a regime characterized by non-linear response sets in, as seen for instance from the red and purple curves. Notice that a 20\% increase of the miRNA transcription rate from the alternative locus can suffice to deviate from a linear behaviour.

Therefore, a sufficiently fast miRNA-ceRNA binding kinetics may lead to a strong variation of the target level even for a modest change in the perturbation size, suggesting an energetically cheap mechanism for the regulation of the expression of a gene involved in a differentiation process, as $m$, that has to be turned on at a definite time in the differentiation program.


\section{Discussion}

We have defined and studied a minimal, deterministic mathematical model of the regulatory circuit that has been recently found to control the timing and expression levels in the early phase of myogenesis \cite{legnini2014feedforward}. We aimed at understanding the roles played by the various mechanisms that appear to coordinate the expression of a target mRNA, including a HuR-controlled channel for the mutually exclusive biosynthesis of the target repressor miR-133 and of its sponge linc-MD1, and an externally-controlled alternative channel for the synthesis of miR-133 from independent genomic loci. In summary, our results indicate that HuR, by controlling the linc-MD1/miR-133 switch, provides essential fine tuning of target levels, while the rapid increase of miRNA transcription from the alternative locus contributes to the fast downregulation of target levels when myoblast differentiation is accomplished. The former mechanism exploits the ceRNA effect and competition to bind miRNAs to indirectly regulate target levels. This strategy has been shown to deliver optimal results at stationarity  \cite{figliuzzi2013micrornas,martirosyan2016probing}. miRNA upregulation has instead been found to be the optimal mechanism to rapidly decrease the level of a target \cite{figliuzzi2014rna}. 

Note that the selection of a miRNA-decoy system over a simpler miRNA-sponge pair, in which the sponge's biosynthesis is unrelated to that of the miRNA, guarantees an optimal level of complementarity (hence a strong coupling) between the miRNA and its lncRNA interaction partner. As we have seen, this condition is crucial for the emergence of non-linear response to a rapid increase of transcriptional activity from the alternative locus. Therefore the presence of a miRNA-decoy system in this context ultimately provides a major advantage both in terms of performance and in terms of metabolic costs.

Revealingly, while static control could be achieved in absence of the alternative channel for miRNA transcription, HuR-based regulation alone would not be as efficient in downregulating the target, as that would require rapid degradation of the controller (i.e. of HuR) to shift miRNA levels up. Both control routes are therefore essential. In addition, the activation of the alternative transcriptional loci can lead to non-linear (negative) response in target levels already for a modest (ca. 20 \%) increase of miRNA biosynthetic rates, providing further support to the idea that miRNAs are optimal dynamical down-regulators.

Our analysis of the complex circuitry controlling skeletal muscle-cell differentiation has been highly simplified on several fronts. In first place, we have focused on a deterministic model, ignoring intrinsic noise sources altogether. The central reason for this lies in the fact that the ceRNA effect as quantified by susceptibilities like (\ref{chimh}) occurs even in absence of fluctuations. Most importantly, in such conditions competition is the only possible source of ceRNA crosstalk (if direct transcriptional dependencies are excluded, as in the present case), as correlations due to the fact that both ceRNAs may respond to stochastic fluctuations in miRNA levels (which do not necessarily signal a positive effective coupling between ceRNAs) are forced to be absent from the picture. In this sense, deterministic models like the one discussed here are ideal to evaluate the role of competition in miRNA-based regulatory elements. Extending the present analysis to include intrinsic (molecular) noise will however allow to quantify the effectiveness of control more precisely by employing different measures, including correlation coefficients \cite{bosia2013modelling} and information-theoretical capacities \cite{martirosyan2016probing}. 

Likewise, it would be important to perform an exploration of parameter space (especially for miRNA-ceRNA interaction constants and complex processing rates) so as to shed light on how robust our conclusions are against kinetic changes. miRNA-based regulatory elements have indeed been shown to be able to exploit kinetic heterogeneities in order to further optimize the efficiency of control with respect to kinetically homogeneous systems \cite{martirosyan2016probing}. While we have focused on the more conservative scenario here, further work in this direction would highlight which functional aspect would most benefit from an accurate tuning of parameters and how the circuit's overall functionality (which encodes for a key developmental program) can be affected by specific perturbations.

\red{Finally, our model has focused for simplicity on a reduced version of the experimentally characterized system, namely one in which a miRNA (miR-135) and a ceRNA (MEF2C) have been suppressed. By influencing the level of linc-MD1, miR-135 however indirectly affects the feedforward loop controlled by HuR. In a full-fledged model accounting for all molecular species shown in Fig. \ref{fig:CIRC}A, then, one should observe a modulation of the levels of linc-MD1 due to its increased sponging activity. As MEF2C is effectively controlled only by miR-135 and, indirectly, by linc-MD1, a down-regulation of the latter will suffice to repress both MEF2C and MAML1, which indeed have similar time courses \cite{legnini2014feedforward}. Therefore, it is reasonable to expect that the overall scenario will be qualitatively identical to that discussed here, albeit with a quantitative shift of the relevant parameter ranges. In specific, changes in $\alpha$ (the probability with which the HuR-controlled switch leads to linc-MD1) and/or $b$ (the intrinsic synthesis rate of linc-MD1) might suffice account effectively for the suppressed molecular species. A more detailed study (involving more parameters) will provide the complete picture.}

\section*{Acknowledgments}

Work supported by the European Union's Horizon 2020 research and innovation programme MSCA-RISE-2016 under grant agreement No 734439 INFERNET.

\section*{References}
\bibliographystyle{elsarticle-num}
\bibliography{biblio}

\begin{thebibliography}{10}
\expandafter\ifx\csname url\endcsname\relax
  \def\url#1{\texttt{#1}}\fi
\expandafter\ifx\csname urlprefix\endcsname\relax\def\urlprefix{URL }\fi
\expandafter\ifx\csname href\endcsname\relax
  \def\href#1#2{#2} \def\path#1{#1}\fi

\bibitem{cech2014noncoding}
T.~R. Cech, J.~A. Steitz, The noncoding {RNA} revolution--trashing old rules to
  forge new ones, Cell 157~(1) (2014) 77--94.

\bibitem{hammond2000rna}
S.~M. Hammond, E.~Bernstein, D.~Beach, G.~J. Hannon, An {RNA}-directed nuclease
  mediates post-transcriptional gene silencing in {D}rosophila cells, Nature
  404~(6775) (2000) 293--296.

\bibitem{bartel2004micrornas}
D.~P. Bartel, Micro{RNA}s: genomics, biogenesis, mechanism, and function, Cell
  116~(2) (2004) 281--297.

\bibitem{bartel2009micrornas}
D.~P. Bartel, Micro{RNA}s: target recognition and regulatory functions, Cell
  136~(2) (2009) 215--233.

\bibitem{kim2016general}
D.~Kim, Y.~M. Sung, J.~Park, S.~Kim, J.~Kim, J.~Park, H.~Ha, J.~Y. Bae, S.~Kim,
  D.~Baek, General rules for functional micro{RNA} targeting, Nature Genetics
  48~(12) (2016) 1517--1526.

\bibitem{ponting2009evolution}
C.~P. Ponting, P.~L. Oliver, W.~Reik, Evolution and functions of long noncoding
  {RNA}s, Cell 136~(4) (2009) 629--641.

\bibitem{rinn2012genome}
J.~L. Rinn, H.~Y. Chang, Genome regulation by long noncoding {RNA}s, Annual
  Review of Biochemistry 81 (2012) 145--166.

\bibitem{fatica2014long}
A.~Fatica, I.~Bozzoni, Long non-coding {RNA}s: new players in cell
  differentiation and development, Nature Reviews Genetics 15~(1) (2014) 7--21.

\bibitem{engreitz2016long}
J.~M. Engreitz, N.~Ollikainen, M.~Guttman, Long non-coding {RNA}s: spatial
  amplifiers that control nuclear structure and gene expression, Nature Reviews
  Molecular Cell Biology 17 (2016) 756--770.

\bibitem{siciliano2013mirnas}
V.~Siciliano, I.~Garzilli, C.~Fracassi, S.~Criscuolo, S.~Ventre,
  D.~di~Bernardo, Mi{RNA}s confer phenotypic robustness to gene networks by
  suppressing biological noise, Nature Communications 4 (2013) 2364.

\bibitem{cheng2007microrna}
H.-Y.~M. Cheng, J.~W. Papp, O.~Varlamova, H.~Dziema, B.~Russell, J.~P. Curfman,
  T.~Nakazawa, K.~Shimizu, H.~Okamura, S.~Impey, et~al., micro{RNA} modulation
  of circadian-clock period and entrainment, Neuron 54~(5) (2007) 813--829.

\bibitem{sayed2011micrornas}
D.~Sayed, M.~Abdellatif, Micro{RNA}s in development and disease, Physiological
  reviews 91~(3) (2011) 827--887.

\bibitem{lai2016understanding}
X.~Lai, O.~Wolkenhauer, J.~Vera, Understanding micro{RNA}-mediated gene
  regulatory networks through mathematical modelling, Nucleic Acids Research 44
  (2016) 6019--6035.

\bibitem{osella2011role}
M.~Osella, C.~Bosia, D.~Cor{\'a}, M.~Caselle, The role of incoherent
  micro{RNA}-mediated feedforward loops in noise buffering, PLoS Computational
  Biology 7~(3) (2011) e1001101.

\bibitem{martirosyan2016probing}
A.~Martirosyan, M.~Figliuzzi, E.~Marinari, A.~De~Martino, Probing the limits to
  micro{RNA}-mediated control of gene expression, PLoS Computational Biology
  12~(1) (2016) e1004715.

\bibitem{marti2}
A.~Martirosyan, A.~{De Martino}, A.~Pagnani, E.~Marinari, ce{RNA} crosstalk
  stabilizes protein expression and affects the correlation pattern of
  interacting proteins, Scientific Reports 7 (2017) 43673.

\bibitem{jens2015competition}
M.~Jens, N.~Rajewsky, Competition between target sites of regulators shapes
  post-transcriptional gene regulation, Nature Reviews Genetics 16~(2) (2015)
  113--126.

\bibitem{salmena2011cerna}
L.~Salmena, L.~Poliseno, Y.~Tay, L.~Kats, P.~P. Pandolfi, A ce{RNA} hypothesis:
  The rosetta stone of a hidden {RNA} language?, Cell 146~(3) (2011) 353--358.

\bibitem{figliuzzi2013micrornas}
M.~Figliuzzi, E.~Marinari, A.~De~Martino, Micro{RNA}s as a selective channel of
  communication between competing {RNA}s: a steady-state theory, Biophysical
  journal 104~(5) (2013) 1203--1213.

\bibitem{bosia2013modelling}
C.~Bosia, A.~Pagnani, R.~Zecchina, Modelling competing endogenous {RNA}
  networks, PLoS One 8~(6) (2013) e66609.

\bibitem{noorbakhsh2013intrinsic}
J.~Noorbakhsh, A.~H. Lang, P.~Mehta, Intrinsic noise of micro{RNA}-regulated
  genes and the ce{RNA} hypothesis, PLoS One 8~(8) (2013) e72676.

\bibitem{figliuzzi2014rna}
M.~Figliuzzi, A.~De~Martino, E.~Marinari, {RNA}-based regulation: dynamics and
  response to perturbations of competing {RNA}s, Biophysical journal 107~(4)
  (2014) 1011--1022.

\bibitem{bosson2014endogenous}
A.~D. Bosson, J.~R. Zamudio, P.~A. Sharp, Endogenous mi{RNA} and target
  concentrations determine susceptibility to potential ce{RNA} competition,
  Molecular cell 56~(3) (2014) 347--359.

\bibitem{yuan2015model}
Y.~Yuan, B.~Liu, P.~Xie, M.~Q. Zhang, Y.~Li, Z.~Xie, X.~Wang, Model-guided
  quantitative analysis of micro{RNA}-mediated regulation on competing
  endogenous {RNA}s using a synthetic gene circuit, Proceedings of the National
  Academy of Sciences 112~(10) (2015) 3158--3163.

\bibitem{denzler2016impact}
R.~Denzler, S.~E. McGeary, A.~C. Title, V.~Agarwal, D.~P. Bartel, M.~Stoffel,
  Impact of {M}icro{RNA} levels, {T}arget-{S}ite {C}omplementarity, and
  {C}ooperativity on {C}ompeting {E}ndogenous {RNA}-{R}egulated {G}ene
  {E}xpression, Molecular cell 64~(3) (2016) 565--579.

\bibitem{poliseno2010coding}
L.~Poliseno, L.~Salmena, J.~Zhang, B.~Carver, W.~J. Haveman, P.~P. Pandolfi, A
  coding-independent function of gene and pseudogene m{RNA}s regulates tumour
  biology, Nature 465~(7301) (2010) 1033--1038.

\bibitem{wang2010creb}
J.~Wang, X.~Liu, H.~Wu, P.~Ni, Z.~Gu, Y.~Qiao, N.~Chen, F.~Sun, Q.~Fan, {CREB}
  up-regulates long non-coding {RNA}, {HULC} expression through interaction
  with micro{RNA}-372 in liver cancer, Nucleic Acids Research 38~(16) (2010)
  5366--5383.

\bibitem{karreth2015braf}
F.~A. Karreth, M.~Reschke, A.~Ruocco, C.~Ng, B.~Chapuy, V.~L{\'e}opold,
  M.~Sjoberg, T.~M. Keane, A.~Verma, U.~Ala, et~al., The {BRAF} pseudogene
  functions as a competitive endogenous {RNA} and induces lymphoma in vivo,
  Cell 161~(2) (2015) 319--332.

\bibitem{cesana2011long}
M.~Cesana, D.~Cacchiarelli, I.~Legnini, T.~Santini, O.~Sthandier, M.~Chinappi,
  A.~Tramontano, I.~Bozzoni, A long noncoding {RNA} controls muscle
  differentiation by functioning as a competing endogenous {RNA}, Cell 147~(2)
  (2011) 358--369.

\bibitem{neguembor2014long}
M.~V. Neguembor, M.~Jothi, D.~Gabellini, Long noncoding {RNA}s, emerging
  players in muscle differentiation and disease, Skeletal muscle 4~(1) (2014)
  8.

\bibitem{legnini2014feedforward}
I.~Legnini, M.~Morlando, A.~Mangiavacchi, A.~Fatica, I.~Bozzoni, A feedforward
  regulatory loop between {HuR} and the long noncoding {RNA} linc-{MD1}
  controls early phases of myogenesis, Molecular cell 53~(3) (2014) 506--514.

\bibitem{shen2006notch}
H.~Shen, A.~S. McElhinny, Y.~Cao, P.~Gao, J.~Liu, R.~Bronson, J.~D. Griffin,
  L.~Wu, The {N}otch coactivator, {MAML1}, functions as a novel coactivator for
  {MEF2C}-mediated transcription and is required for normal myogenesis, Genes
  \& development 20~(6) (2006) 675--688.

\bibitem{valencia2006control}
M.~A. Valencia-Sanchez, J.~Liu, G.~J. Hannon, R.~Parker, Control of translation
  and m{RNA} degradation by mi{RNA}s and si{RNA}s, Genes \& development 20~(5)
  (2006) 515--524.

\bibitem{wang2010toward}
X.~Wang, Y.~Li, X.~Xu, Y.-h. Wang, Toward a system-level understanding of
  micro{RNA} pathway via mathematical modeling, Biosystems 100~(1) (2010)
  31--38.

\bibitem{haley2004kinetic}
B.~Haley, P.~D. Zamore, Kinetic analysis of the {RNA}i enzyme complex, Nature
  Structural \& Molecular Biology 11~(7) (2004) 599--606.

\bibitem{alon2006introduction}
U.~Alon, An introduction to systems biology: design principles of biological
  circuits, CRC press, 2006.

\bibitem{palssonsystems}
B.~Palsson, Systems biology: properties of reconstructed networks. 2006,
  Cambridge Univ Press.

\bibitem{translating}
A.~Martirosyan, M.~Marsili, A.~{De Martino}, Translating ce{RNA}
  susceptibilities into correlation functions, bioRxiv preprint
  [http://www.biorxiv.org/content/early/2017/01/25/102988].

\end{thebibliography}

\end{document}